\documentclass[showpacs,twocolumn,aps,prb,byrevtex]{revtex4}
\usepackage{graphicx}
\begin{document}

\title{
Theoretical investigations of a highly mismatched interface: the case of
 SiC/Si(001)
}
\author{Laurent~Pizzagalli}
   \email{Laurent.Pizzagalli@univ-poitiers.fr}
\affiliation{Laboratoire de M{\'e}tallurgie Physique, UMR 6630, CNRS \&
Universit{\'e} de Poitiers, BP 30179, F-86962 Futuroscope Chasseneuil
       Cedex, France}
\author{Giancarlo~Cicero}
\affiliation{CNR-IMEM, Parco Area delle Scienze, 37A, I-43010
Parma, Italy; and INFM and Physics Department, Polytechnic of
Torino, C.Duca degli Abruzzi, 24,
       I-10129 Torino, Italy}
\author{Alessandra~Catellani}%
\affiliation{CNR-IMEM, Parco Area delle Scienze, 37A, I-43010
Parma, Italy }
\date{\today}

\begin{abstract}

Using first principles, classical potentials and elasticity
theory, we investigated the structure of a
semiconductor/semiconductor interface with a high lattice
mismatch, SiC/Si(001). Among several tested possible
configurations, a heterostructure with (i) a misfit dislocation
network pinned at the interface and (ii) reconstructed dislocation
cores with a carbon sub-stoichiometry is found to be the most
stable one. The importance of the slab approximation in first
principles calculations is discussed and estimated by combining
classical potential techniques and elasticity theory. For the most
stable configuration, an estimate of the interface energy is
given. Finally, the electronic structure is investigated and
discussed in relation with the dislocation array structure.
Interface states, localized in the heterostructure gap and located
on dislocation cores, are identified.

\end{abstract}

\pacs{68.35-p, 81.05.Hd, 81.15.Aa}

\maketitle

\section{Introduction}

The misfit strain present in lattice mismatched epitaxial layers
has been widely studied because of its omnipresence in the
heterostructures used in device technology. Two goals are usually
pursued, either to avoid strain for preparing long-lived devices,
or to exploit the modified electronic properties of strained
layers for obtaining specific devices such as 
lasers.\cite{Dun97JMSME,Jai97PMA} The general picture of heteroepitaxy
is well known. First, strained layers are grown on the substrate.
Over a critical thickness, which depends on the elastic properties
and the lattice mismatch between the two materials, plastic
relaxation of the strain with formation of misfit dislocations
becomes energetically favorable. In the case of small mismatch,
the critical thickness can be large, with a low dislocation
density. For example, a critical thickness of about $10^3$~\AA\ is
measured for a lattice mismatch of 1\% in Ge$_x$Si$_{1-x}$ layers
($x = 0.24$) \cite{Bea92PIEEE}, and the average separation between
misfit edge dislocations is estimated to be about 390~\AA\ (for
[001] layers). Each dislocation is far from the others and from
the interface, and the system can be described by considering an
ideal coherent interface, with strained layers. Hoekstra and
Kohyama used this frame for investigating the $\beta$-SiC(001)/Al
interface.\cite{Hoe98PRB} In a scheme recently proposed, Benedek
{\sl et al.} introduced an additional correction for the effect of
the misfit dislocations \cite{Ben02JPCM}, by comparing
coherent and semi-coherent interfaces. However, in the case of
heteroepitaxy for largely mismatched systems, the critical
thickness is very low and a dense network of misfit dislocations
is present in the grown layers. The interactions between
dislocations and between the interface and the dislocations must
then be explicitly taken into account.

Despite the large lattice mismatch of $\sim$20$\%$, $\beta$-SiC
can be grown on Si(001),  using different 
techniques.\cite{Nut87APL,Wah94APL,Lon99JAP,Kit00TSF} A recent
High-Resolution Electron Microscopy (HREM) study of this system
shows a locally abrupt interface, with the presence of a periodic
array of misfit dislocations.\cite{Wah94APL,Lon99JAP,Kit00TSF}
These ones seem to be located directly at the interface, which
points at a vanishing critical thickness for this heterostructure.
It is difficult to gain additional information from these
experiments. In particular, the atomic structure and the chemical
environment at the interface, which deeply affect its physical
properties, still remain hardly accessible. Atomistic computations
are a possible solution for complementing the experiments. In
addition, first principle calculations would allow the
determination of the electronic properties of the heterostructure.
In this respect, the SiC/Si interface may be considered as a model
for a high lattice mismatch interface between covalent materials,
just as silicon is usually considered as the semiconductor
prototype. Indeed, due to the peculiar lattice mismatch, almost
equal to 20\%, the network dislocation pattern is extremely dense
and can be modelled within a cell small enough to make an ab
initio computation affordable.

As far as we know, few studies have been devoted to the
SiC/Si(001) system. Chirita {\sl et al.}~\cite{Chi97TSF} have
investigated the strain relaxation and the thermal stability of
the interface using the semi-empirical classical potential from
Tersoff.\cite{Ter89PRB} They have obtained a possible
configuration of the atomic structure, stable up to 1000~K.
However, they assumed a stoichiometric interface, which may be a
metastable state. Another study from Kitabatake with the same
potential choice was exclusively focussed on the first step of the
formation of the interface, {\sl i.e.} a pre-carbonization of the
Si(001) surface prior to growth.\cite{Kit00TSF} A detailed
investigation of the atomic structure and stoichiometry of the
SiC/Si(001) interface with first principle techniques is still
lacking. The large computational cell required for the modelling,
notwithstanding the advantageous lattice mismatch, may explain the
absence of such calculations.

Recently, we have presented in a letter the most striking results
of the first ab initio study of the SiC/Si(001) interface.\cite{Cic02PRL}
 Here, we provide a thorough description of our
investigations, combining elasticity theory, classical potentials
and first principles calculations. In particular, after a complete
explanation of the computational procedure, the full set of tested
atomic configurations is presented. We have investigated the
stoichiometry of the interface as well as its stability with few
pseudomorphic SiC layers. The determination of the interface
energy as well as the analysis of finite size effects in the slab
approximation are also presented. Finally, the computed electronic
structure of the most stable configuration is discussed.

\section{Computational procedure}

\subsection{Model}

Considering that the lattice parameters of Si and 3C-SiC are
respectively $a(\mathrm{Si})=5.4309$~\AA\ and
$a(\mathrm{SiC})=4.3596$~\AA\ \cite{SiClatt}, the calculated
misfit
$\left(a(\mathrm{Si})-a(\mathrm{SiC})\right)/a(\mathrm{Si})$ at
the interface is 19.73\%. Within the hypothesis of a coherent
interface, such a misfit is equivalent to a huge tensile stress in
the SiC layers. A simple estimate shows that the elastic energy
stored in the coherent film would be approximately
1.6~J/m$^{2}$/layer.\cite{cohelast} However, it is observed that the interface is
semi-coherent, the stress being relaxed by the introduction of a
network of undissociated edge dislocations of Burgers vector
$b=a(\mathrm{SiC})/2\langle110\rangle$, with dislocation lines
lying along the $[110]$ and $[\bar{1}10]$ 
directions.\cite{Wah94APL,Lon99JAP,Kit00TSF} Due to the peculiar value of
the misfit, approximately equal to $\frac{1}{5}$, the
semi-coherence is obtained when the spacing between misfit
dislocations is $5b$. The SiC/Si(001) interface is then modelled
by matching a p($5\times5$)-SiC(001) slab (N layers) with a
p(4$\times4$)-Si(001) slab (M layers) along the $[001]$ direction
(z-axis), at the Si lattice parameter, optimized for bulk
calculations with the chosen method. In the following, size
conditions for a system will be simply designed by N/M. Periodic
boundary conditions are applied along $[110]$ and $[\bar{1}10]$
directions (x- and y-axis) in any case. Atoms belonging to the
topmost SiC and bottommost Si layers were fixed in bulk positions.
However, in order to better relax the interface, the SiC topmost
layer was allowed to move as a whole in all directions. This
procedure was achieved at each step by (i) compute the average of
atomic forces in this layer, (ii) apply this force to all layer
atoms. Once the structures were converged, we also performed
additional calculations to check the effect of the surfaces on the
interface by releasing the constraints on both surfaces.

\subsection{Methods}

Classical molecular dynamics calculations were done with the
semi-empirical Tersoff potentials \cite{Ter89PRB}, at the lattice
parameter $a(\mathrm{Si})=5.43$~\AA. This choice is currently the
best available for describing solid SiC. A recent study has
revealed that it reproduces DFT results on the formation energies
and properties of native defects in SiC with a good 
accuracy.\cite{Gao01NIMPR} Here, several system sizes were considered,
because of the low CPU time and memory requirements for these
calculations. The exploration of the configurational space was
typically done with 12/8 and 36/36 slabs, to ensure that the
interaction between the interface and the surfaces is negligible.
To estimate this interaction, sizes as large as 50/50 have been
considered.

First principles calculations were performed \cite{BASICJEEP}
within the Density Functional Theory (DFT), in the Local Density
Approximation (LDA).\cite{Koh65PR} An energy cutoff of 40 (160)
Ry was used for the plane waves expansion of the wavefunctions
(charge density). The reciprocal space integration in the
supercell Brillouin Zone (BZ) was done by considering only the
$\Gamma$ point. The core electrons were removed by using
pseudopotentials, with {\sl s} and {\sl p} non-locality for Si and
{\sl s} non-locality for C.\cite{Ham89PRB} We used a calculated
Si lattice parameter of 5.401~\AA, consistent with our ab initio
pseudopotential. Surface atoms at both sides were saturated with
hydrogens. The C-H and Si-H distances were optimized independently
by fully relaxing surface saturated symmetric slabs of 11 layers,
8 atoms each. 5/5 and 7/7 system sizes were used, which
corresponds to large scale calculations with at most 369 atoms. At
variance with classical simulations, periodic boundary conditions
were also applied along the $[001]$ direction, a large vacuum
space (9.1~\AA\ for 5/5, 8.2~\AA\ for 7/7) being introduced to
prevent spurious surface-surface interactions. All structures were
considered converged when forces acting on atoms were less than
10$^{-4}$ a.u. (0.005 eV/\AA) and energy varied by less than
10$^{-5}$ eV/atom.

\subsection{Configuration energy}

In this work, we have always compared the energies of systems with
the same N/M. The numbers of C and Si atoms in the SiC slab
may however change for two N/M slabs with different
interface configurations. In addition, the surface of the SiC slab
may be either Si- or C-terminated. Hence, the energy $E_{slab}$ of
the N/M slab model described above, obtained from either methods,
cannot be used directly for determining the most stable structure.
Instead, a configuration energy $E_\alpha$, given by the following expression,
can be defined within the grand canonical frame:

\begin{equation}
E_\alpha=E_{slab}-E_S^\mathrm{SiC}-E_S^\mathrm{Si}-n^\mathrm{Si}\mu^\mathrm{Si}-n^\mathrm{C}\mu^\mathrm{C}
\end{equation}

\noindent $E_S^\mathrm{SiC}$ is the surface energy of the SiC part
of the slab, Si- or C-terminated depending on the configuration at
the interface and on the number of layers, whereas
$E_S^\mathrm{Si}$ is the surface energy of the Si part.
$n^\mathrm{Si}$ and $n^\mathrm{C}$ are the numbers of Si or C
atoms in the SiC/Si slab. Finally, $\mu^\mathrm{Si}$ and
$\mu^\mathrm{C}$ are the chemical potentials for each species. For
the Si part of the slab, $\mu^\mathrm{Si}$ is equal to the bulk
silicon chemical potential.
Though $\mu^\mathrm{Si}$ and
$\mu^\mathrm{C}$ cannot be exactly determined for the SiC part, it
is possible to determine a range of thermodynamically allowed
values.\cite{Qia88PRB} The following relations must be satisfied
at the same time:

\begin{eqnarray}
\mu^\mathrm{Si}+\mu^\mathrm{C} & = & \mu_0^\mathrm{SiC} \\
\mu_0^\mathrm{Si}+\mu_0^\mathrm{C} - \Delta\mathcal{H}\!_f & = &
\mu_0^\mathrm{SiC} \label{chempot10}
\end{eqnarray}

\noindent Here, the $\mu_0^I$ are the chemical potentials of the
mono-elemental bulk phases, and are easily calculated.
$\Delta\mathcal{H}\!_f$ is the heat of formation of silicon
carbide: in this work, we used the experimental value,
$\Delta\mathcal{H}\!_f$ = 0.72 eV. These two relations reveal
that the Si and C chemical potential values are linked, and that
the chemical potential of one species is always equal or lower
than the chemical potential of the mono-elemental bulk phase. The
allowed ranges are then

\begin{eqnarray}
\mu^\mathrm{Si}&\in &\left[\mu_0^\mathrm{Si}-\Delta\mathcal{H}\!_f,\mu_0^\mathrm{Si}\right]\\
\mu^\mathrm{C}&\in
&\left[\mu_0^\mathrm{C}-\Delta\mathcal{H}\!_f,\mu_0^\mathrm{C}\right]
\end{eqnarray}

\noindent
$(\mu_0^\mathrm{Si}-\Delta\mathcal{H}\!_f,\mu_0^\mathrm{C})$
corresponds to C-rich (Si-poor) conditions and
$(\mu_0^\mathrm{Si},\mu_0^\mathrm{C}-\Delta\mathcal{H}\!_f)$  to
C-poor (Si-rich) conditions. The values obtained in the present
work for $\mu_0^I$ are listed in Table~\ref{chempot}: here, for
the ab initio results, we indicate the respective values as
obtained from standard converged bulk calculations and as
extrapolated from surface calculations with the scheme originally
proposed by Fiorentini and Methfessel.\cite{Fio96JPCM} This will
be useful to determine the different contributions in Eqn.~(1) and
the interface energy, as described below. It is worth noting that
extreme caution must be adopted when estimating $\mu_0^I$, as
small ($\simeq 1\%$) errors in this quantity are then multiplied
by the usually large number of atoms at the interface, eventually
producing non-physical results. The use of the linearized values
of $\mu_0^I$ in this work yields different configuration energy
values with respect to our previously published results
\cite{Cic02PRL}, without changing the original qualitative
conclusions.

Using the configuration energies allows to compare interface
structures in the slab approximation, with different surface
terminations and number of atoms. In the following, we discuss the
stability of different configurations.

\section{Stability and geometry}

In a previous work on the SiC/Si(001) interface, Chirita {\sl et
al.} proposed a possible geometry of the interface
\cite{Chi97TSF}, based on molecular dynamics calculations. In the
choice of their initial atomic configuration, they have made
several hypothesis. They assumed that the dislocations network is
pinned directly at the interface, that the first (001) layer of
the SiC slab is silicon-like, and finally, that the interface
remains perfectly stoichiometric. However, from HREM experiments,
it is hard to extract such information. An interface with a
pseudomorphic first layer followed by a misfit dislocations
network is possible, as well as reconstructed misfit dislocation
cores including a sub- or over-stoichiometry in carbon and/or
silicon atoms. In our work, we have taken into account such
possibilities, and our search for the most stable configuration
has been broadened by considering a large set of initial
geometries, by means of classical molecular dynamics. First, only
interfaces made with complete layers were investigated.
Fig.~\ref{config1} shows 6~different possible starting geometries.
The first ones, S1 and S2, are stoichiometric configurations with
the misfit dislocation array located directly at the interface. In
S1, the first SiC layer is carbon-like whereas it is silicon-like
in S2 (S2 is the initial configuration selected in the previous
study \cite{Chi97TSF}). In P1 and P2, the misfit dislocation array
is located in the second and third SiC layers, respectively. Those
are obtained by removing one or two p(5$\times5$) SiC layers,
replaced by p(4$\times4$) pseudomorphic layers. Finally, we also
compared the previous structures with configurations C1 and C2,
with a higher carbon concentration at the interface
(carbonization).

The configuration energy differences
are reported in Table~\ref{Energy1}. Here, the
calculated values are easily extracted from the classical
dynamics, as the energy of each atom is a well defined quantity in
the interaction potential model. The S2 configuration,
investigated in a previous study \cite{Chi97TSF}, is stable but is
clearly not the lowest energy solution. Instead, the most stable
structures are obtained for a stoichiometric system and a C layer
at the interface, {\sl i.e.} S1. For such structures, we found two
different atomic configurations, represented in
Fig.~\ref{configrelax1}. It must be emphasized that the two
$[110]$ and $[1\bar{1}0]$ directions, perpendicular to the
interface, are not equivalent due to the peculiar zincblende
stacking. The first configuration, called S1a, has the lowest
energy in the full range of chemical potentials. Along $[110]$,
bonds are formed between the extra C atoms located at the misfit
dislocation core and Si atoms belonging to the second layer of the
Si(001) slab. All these atoms are shifted toward the interface
inducing large strains, especially in the softer silicon slab.
Some of the Si atoms appear to be overcoordinated. Along
$[1\bar{1}0]$, the main feature is a soft dimerization of the C
atoms of the first SiC layer, on both sides of the misfit
dislocation line. The C-C distance is shortened to 1.65-1.80~\AA\
which allows a reduction of the number of dangling bonds (DB).
This interface reconstruction is nearly similar to the model
presented by Long {\sl et al.}.\cite{Lon99JAP} The second
configuration, S1b, has a higher configuration energy, with
6.55~eV in excess. The main difference with S1a concerns the
$[110]$ side. Here, no bonds are formed between the C atoms in the
first SiC layer and the second layer of the silicon slab. Instead,
the number of DBs belonging to C atoms is reduced by a stronger
carbon dimerization along the $[110]$ direction (clearly visible
on the $[1\bar{1}0]$ view), where the C-C distance is reduced to
1.44-1.47~\AA. From Table~\ref{Energy1}, it seems that the
occurrence of pseudomorphic SiC layers before the introduction of
misfit dislocations is not favored, although the P1 structure,
with one pseudomorphic layer, is the third best solution. The
segregation of carbon at the interface, tested with two
configurations C1 and C2, is associated with very high
configuration energies. This is consistent with the general
observation that the deposition of carbon on Si(001) surface does
not lead to stable carbon layers but rather to the formation of
silicon carbide by carbonization, the substrate supplying the Si
atoms.\cite{Kit00TSF}

Several information may be obtained from our results. First, one
or several complete pseudomorphic SiC layers are not favorable
structures, which tends to indicate that the misfit dislocation
network is located at the interface and no finite critical
thickness can be defined. Secondly, all the favored configurations
include a carbon layer at the interface. This can be explained by
the energy gained in the formation of SiC bonds, and the smaller
carbon atomic radius with respect to silicon. Finally, we observed
that the minimization of the configuration energy is best realized
by the reduction of DBs, especially those associated to carbon
atoms. However, this DB saturation occurs via the formation of
bonds between initially remote atoms, and the shortening of their
separation generates energetically expensive strain in the atomic
structure. This is particularly true for the shortest and strongest
carbon-carbon or silicon-carbon bonds.

For the lowest energy configurations, S1a and S1b, the carbon
atoms, located in the first SiC layer and in the vicinity of the
misfit dislocation cores, present the most stretched bonds. We
investigated the possibility of an energy lowering by removing
these atoms. We initially considered the S1b configuration. A
significant energy reduction was obtained by removing successively
C atoms along $[110]$ (inside the ellipsis, in
Fig.~\ref{configrelax1}). The additional removing of the C-dimer
row along $[1\bar{1}0]$ (the other ellipsis) leads to an even more
stable configuration, represented in Fig.~\ref{configrelax2}. The
energy difference compared to S1a is $-6.04$~eV/cell
($-15.40$~eV/cell) in C-rich (C-poor) conditions, 
respectively.\cite{notePRL} Starting from S1a, and removing the C-row along
$[110]$ and the C-dimer row along $[1\bar{1}0]$ as described
previously, the energy was also lowered, and we obtained the same
final structure. This configuration, which we called CSS for
Carbon SubStoichiometric, shows remarkable features. Along the
$[110]$ direction, dangling bonds created by the removal of C
atoms are eliminated with the formation of Si dimers
2.48-2.54~\AA\ long. Compared to S1a, the Si atoms of the silicon
slab, previously bonded to C atoms, recover bulk-like positions,
thus minimizing the strain. Along $[1\bar{1}0]$, after the removal
of the C-dimer row, the core of the misfit dislocation is made of
8-atoms rings, including seven Si atoms and a lone C atom. Almost
all these atoms are fully coordinated, at the expenses of some
bond stretching, particularly on the Si atoms in the second layer
of the SiC slab, and at the intersection of the two perpendicular
dislocations. The Si-Si bonds range from 2.33 to 2.62~\AA. We
found that CSS is the most stable configuration using the Tersoff
potential. Indeed, tests performed on structures with further C
removal, or selected C/Si exchanges, brought no additional energy
reduction.

The Tersoff potential results were confirmed by the ab initio
method. Owing to the huge computational effort required to deal
with the interface, our investigations were restricted to three
configurations, {\sl i.e.} the low energy stoichiometric S1a and
S1b, and the most stable geometry CSS. First, we found that only
S1a and CSS were stable,  S1b relaxing spontaneously to S1a. It is
known that classical potentials tend to stabilize a larger number
of metastable structures compared to first principles methods.\cite{Bal92PRB,Sbr02SS} No
major geometrical changes were found for both configurations
relaxed with first principles, compared to the classical results.
A better description of the atomic structure of the reconstructed
dislocation cores was however obtained. Considering CSS and the
$[110]$ direction, the Si dimer lengths range between 2.42 and
2.50~\AA. Along $[1\bar{1}0]$, the dislocation core is made of Si
bonds 2.36-2.52~\AA\ long. Thus the ab initio relaxation yields a
more compact 8-atom ring. The energy differences between S1a and
CSS are reported in Table~\ref{Energy2}, for both methods and
different slab lengths. Despite the small sizes imposed by ab
initio simulations, it appears that the energy difference is
already well converged for a 5/5 slab, with a 0.2~eV/cell
variation between 5/5 and 7/7. The ab initio calculation validates
our primary results, {\sl i.e.} the CSS configuration is the most
stable one, for the whole allowed range of the chemical
potentials. Moreover, even in C-rich conditions, the energy
difference between the two geometries is much larger than the
error associated with such computations.

The analysis of the CSS configuration topology gives some hints
for understanding this result. The first set of calculations has
indeed shown that it is preferable to have C atoms at the
interface in the first SiC layer.
From the second set, it appears that misfit dislocation cores with
only silicon atoms are favored, the stretching of carbon bonds
being energetically expensive. In CSS, the first SiC layer is
carbon-like, but there is only one C atom involved in the
reconstructed dislocation cores. It also presents the peculiar
characteristics that almost no atoms are sub- or over-coordinated,
owing to the formation of a topological ring along one direction
and the formation of a silicon dimer along the other. The CSS
configuration is then the best candidate to represent the atomic
structure of the SiC/Si(001) interface.

A qualitative indication of the residual strain distribution at
the interface can be obtained by inspection of the deformation of
the converged system coordinates, with respect to the ideal
bulk-like positions. In Fig.~\ref{warping}, we report the
layer puckering defined as the maximum deviation in the direction
perpendicular to the interface, as obtained from ab initio
calculations for a 7/7 system: the major deformations are
localized in Si, which has smaller elastic constants than SiC. The
warping decreases when moving aside the interface, in agreement
with experimental results \cite{Xu00PRL} on the structural
characterization of SiC films grown on a Si(001) substrate, which
evidenced an internal roughness of individual SiC planes that
diminishes away from the silicon substrate. Fig.~\ref{stress}
represents a comparison of the residual strain field at the
interface for both the CSS and S1a geometries, evaluated in terms
of atomic displacements from ideal bulk-like positions. The
superior efficiency of CSS in relieving the strain is clearly
evidenced by simple inspection.

Regarding the critical thickness, it is interesting to
compare the CSS and P1 configurations. Indeed, the SiC film in the
CSS geometry is constituted by a substoichiometric C layer, as
if obtained from P1 by removing four C atoms in a row. It is thus
difficult to define precisely a finite critical thickness for this
interface, characterized by largely reconstructed dislocation
cores, although our results support the experimental findings of a
dislocation network pinned at the interface.\cite{Lon99JAP}

\section{Effect of slab size}

The large lattice misfit between $\beta$-SiC and Si allows the
investigation of the (001) interface with ab initio methods, the
spacing between dislocations being about
$a(\mathrm{SiC})/2\langle110\rangle \simeq$ 15.4~\AA. There are
only 4 (Si portion) or 5 (SiC part) atoms per $\langle 110\rangle$
edge. However, because the problem is two-dimensional, a slab
layer includes 16 (Si part) or 25 (SiC part) atoms.
Computationally, the number of layers one can use to model the
interface is then severely limited. Here, we managed to calculate
at most a 7/7 interface, {\sl i.e.} $\simeq$ 300 atoms. Such size
is enough when considering a coherent interface. Here, the
presence of a periodic network of misfit edge dislocations at the
interface induces a strain field in both the SiC and Si portions
of the slab. It is usually assumed that the strain field
penetrates each part by a distance of the order of the dislocation
spacing.\cite{Ben02JPCM} In our case, we should then investigate
a 14/11 slab, which is beyond the capabilities of available
supercomputers. Our largest calculations performed on a 7/7 slab
indeed reveal that the strain field is still not negligible at
both ends of the slab: when allowing for a global relaxation, the
two surfaces became slightly bent, due to residual strain, which
may furthermore be different for different core structures. The
flat surface constrain adopted in our simulations is a valuable
approximation to perform energy comparison: this constrain however
modifies the strain field of the dislocation arrays and a
surface-interface interaction is present in the system. This
interaction depends on the core structure at the interface, and
may affect the relative stability of the configurations.

In this part, we investigate the slab size effect by means of
classical potential calculations and elasticity theory. The
configuration energy, defined previously in Eqn.~(1), may be
written as the sum of four contributions:

\begin{equation}
E_\alpha(h)=E_a+E_{c}+E_{el}(h)+E_{is}(h)\label{intersurf}
\end{equation}

\noindent Here, $h$ is the slab size. $E_a$ is a constant adhesive
energy between Si and SiC parts; $E_c$ is the core energy of the
misfit dislocation network; $E_{el}$ is an elastic energy due to
the strain field; $E_{is}$ is the interaction energy between
surface and interface, which is zero for a slab including a large
(infinite) number of layers. Only three terms depend on the slab
size $h$. $E_{el}$ is determined by using isotropic
elasticity theory and a model of misfit dislocation arrays at the
interface between a thin film of height $h$ and an infinite
substrate.\cite{Bon92PMA} The dislocation core radius
\cite{Hir82WIL} is assumed to be equal to the Burgers vector $b$,
{\sl i.e.} the in-plane SiC lattice parameter
$a(\mathrm{SiC})/2\langle110\rangle$, in the present case.
Provided that $E_{el}$ is known for
each $h$, $E_{is}$ may be obtained from Eqn.~\ref{intersurf} by
calculating the configuration energy $E_\alpha$ with increasing
slab size, for a chosen dislocation network. Here, the
surface-interface interaction energy is determined separately for
both the Si and SiC parts of the slab. We performed Tersoff
potential calculations for slabs 31/n (increasing the Si part) and
n/31 (increasing the SiC part), with n ranging from 5 to 31, and
for both the S1a and CSS configurations.

The calculated interaction energies between surface and interface,
$E_{is}$, are shown in Fig.~\ref{surfinternrj}, for both
configurations, in both portions of the slab, together with the
sum of these terms. The SiC contribution appears larger than the
Si one, for a given number of layers. Note that the interlayer
spacing for SiC is about 20\% lower than for Si, and more SiC
layers are needed to get an equivalent contribution from the SiC
and Si parts of the slab. For example, a 10/8 slab will have SiC
and Si parts of nearly identical weights on $E_{is}$. A larger SiC
contribution is then expected for a lower number of layers. Here,
the difference is important, in particular for CSS with an
interaction energy almost three times larger for SiC than for Si.
This could be explained by the geometry of CSS. The dislocation
core along one direction is reconstructed with a ring of atoms,
almost entirely located in the SiC part of the slab.
The surface/interface interaction is then stronger in the SiC
part.

Our results confirm that the strain field penetrates by a distance
of the order of the dislocation spacing \cite{Ben02JPCM}, and that
ideally a 14/11 slab should be used. In fact, the sum of the SiC
contribution for 14 layers with the Si contribution for 11 layers
is less than 0.02~eV/cell, for both solutions. Assuming that the
surface/interface interaction energy is mostly elastic, it is
reasonable to consider that this quantity could be fairly
estimated using classical potentials. Considering the total
interaction energy $E_{is}$ in Fig.~\ref{surfinternrj}, for the
S1a configuration, $E_{is}$ is 0.66~eV/cell (2.36~eV/cell) for a
7/7 (5/5) slab, while for CSS, $E_{is}$ is 0.80~eV/cell
(3.69~eV/cell) for a 7/7 (5/5) slab. In all cases, $E_{is}$ is
lower than the calculated configuration energy differences, and our previous
conclusions on the stability of the CSS configuration remain
valid. Moreover, since the slab size effect is stronger for CSS
than for S1a, using larger slabs will further increase the
stability of the CSS configuration.

\section{Interface energy}

As far as we know, there is no measured value of the SiC/Si(001)
interface energy.  Experimentally, it is possible to determine the
bonding energy, which is related to the interface energy, from
wafer bonding experiments. However, a large range of values may be
obtained, depending on the kind of SiC polytypes or surface
terminations.\cite{Plo99MSE} From the theoretical point of view,
no value is available. Possible explanations are either the large
size of the system that have to be dealt with first principles
methods, or the difficulty to extract such energy from a slab
calculation. A simplified frame for obtaining the interface energy
of mismatched interfaces has been proposed 
recently.\cite{Ben02JPCM} The method however does not take into account a
possible reconstruction or understoichiometry of the core of the
misfit dislocations, which is mandatory for the SiC/Si(001)
system, as discussed above.

In order to extract the interface energy of the CSS configuration,
the energies of the surfaces on both sides of the slabs have to be
known. However, with ab initio methods, only the total energy is accessible, and
surface energies cannot be obtained directly. Instead, they were
determined by extrapolating from slabs with an increasing number
of layers, following the scheme proposed by Fiorentini and
Methfessel.\cite{Fio96JPCM} Si(001)-(1$\times$1) and
C:SiC(001)-(1$\times$1) hydrogenated surfaces were investigated,
since only the carbon terminated surface was relevant for the
selected configurations. In order to minimize computational
errors, a large (c(4$\times$4)) slab, with increasing thickness,
from 7 up to 15 layers, was used: the extrapolated values for the
chemical potentials are close to those obtained from bulk
calculations for Si and SiC respectively (see
Table~\ref{chempot}).
To be consistent with the SiC/Si interface slab calculation, each
surface atom was saturated by 2 symmetric hydrogens. As a
consequence, the surface energies reported in Table~\ref{surfnrj}
include the energies of the pseudo-hydrogens. We found a C-H
distance of 1.09~\AA\ and a bond angle $\widehat{\mathrm{HCH}}$ of
100.6$^\circ$ for the C:SiC(001)-(1$\times$1), and a Si-H distance
of 1.47~\AA\ and a bond angle $\widehat{\mathrm{HSiH}}$ of
101.4$^\circ$ for the Si(001)-(1$\times$1). We point out that we
applied constraints in order to keep a symmetric dihydride
Si(001)-(1x1) surface, although the most stable configuration is
canted.\cite{Nor91PRB} We forced the symmetric geometry to
quickly recover a bulk-like behavior, as required by the small
slab sizes adopted in the calculations, since the correct canted
configuration extends deeply in inner layers.

The interface energy $\sigma_I$ is obtained as the converging
value of the configuration energy for large N/N slabs. In the
classical case, it is easily estimated with large slabs, where
the interaction between surface and interface is negligible, and
all the elastic energy can be considered fully pertinent to the
inner layers of the slab. In this case, we obtain
$\sigma_I\simeq$ 22.7~eV/cell, ({\sl i.e.} $\sigma_I\simeq$
1.6~J/m$^{2}$).

Using ab initio methods, we are limited to small slabs, 5/5 and
7/7. In principle, the determination could be done in a similar
way than for surface energies, using an extrapolating 
technique.\cite{Fio96JPCM} However, here, the available slab sizes are too
small for that purpose. In fact, both elastic and
interface-surface energies considerably change between the 5/5 and
the 7/7 slabs. As a consequence, a linear regression would yield
misleading values for the slope and intercept constants. The
extrapolating scheme could be used for slabs large enough to have
constant elastic and surface-interface interaction energies, {\sl
i.e.} for N/N slabs with N greater than 14 for example. Instead,
we directly calculate the interface energy for a 7/7 slab, using
the chemical potentials obtained via linear extrapolation, as
discussed above.
We computed $\sigma_I\simeq 23.0$ (22.6) eV/cell in C-rich
(C-poor) conditions.
This quantity is not exactly the interface energy. Indeed, for a
7/7 slab, the surface-interface interaction is not negligible and
the strain energy is still not fully converged. These
contributions have been determined in the previous section: for a
7/7 slab in CSS geometry the excess surface-interface interaction
is 0.8~eV/cell (see Fig.~4), and the calculated residual elastic
energy between an infinite interface and a 7/7 slab is
0.2~eV/cell. The corrected interface energy is then
$\sigma_I\simeq$ 22.4 (22.0) eV/cell ({\sl i.e.} 1.58 (1.55)
J/m$^{2}$) in C-rich (C-poor) conditions.
The agreement between classical and ab initio values is surprinsigly good 
and maybe fortuitous given the technical difficulties associated with the
interface energy determination in the ab initio case, or the
use of the Tersoff potential that does not include properly the
electronic contributions.

We wish to stress again at this point that the ab initio values provided for
the interface energy are mere estimates, as small variations in
{\sl e.g.} the chemical potentials can induce large errors and
even non-physical negative interface energies. Our choice in
presenting these results has been to keep consistency between the
chemical potentials used to eliminate the surface and bulk
contributions: for this reason, we always adopted the values for
$\mu_0^I$ as obtained in the linear extrapolation scheme
\cite{Fio96JPCM}, for large supercells, that allowed a good
$k_{\parallel}$ sampling.

\section{Electronic properties}

We now move to the description of the electronic structure of the
mismatched heterostructure. The interface configuration determines the
electronic properties of the system: the presence of defects such
as misfit dislocations can induce interface states in the band
gap, that may severely modify device performances. Indeed, our
results indicate that a number of interface derived states lay in
the forbidden energy gap for the most stable dislocation network,
although the number of DBs is in this case minimized.

For a 5/5 slab, the valence band widths (VBW) of Si- and
SiC-derived bulk states compare fairly well for frozen and free
surfaces, although they underestimate the respective bulk
calculations (Table~\ref{VBW}). Increasing the slab thickness to
7/7 leads to VBW variations of only $\simeq 3\%$. We were able to
estimate an error due to the slab approximation of $\pm$ 0.3 eV on
LDA eigenvalues. A 7/7 slab is then large enough to get a good
description of both the Si and SiC part, and the perturbation
induced on the electronic structure by the interface
configuration. In Fig.~\ref{CSSdos}, the spatially projected
Density Of States (DOS) at the interface is compared with
the DOS obtained in inner layers at the Si and SiC sides of the
slab. Several states lay in the band gap of the heterostructure,
as obtained by alignment of the respective bulk valence bands 
(see Table~\ref{VBW} and the peak
above the valence band top at $\Gamma$, highlighted by the arrow
in Fig.~\ref{CSSdos}). The Highest Occupied (HO) and Lowest
Unoccupied (LU) states are located at 0.7 and 1.1~eV above the
valence band top at $\Gamma$. These two states are also localized
in the core of the edge dislocations, as a result of the large
difference in charge transfer between Si-C and Si-Si bonds. In
Fig.~\ref{figHOMO}, the charge density plot of the HO state along
the two dislocations directions is represented. The charge density
is mainly localized on atoms of the [1$\bar{1}$0] core
dislocation, while no density is observed around the other core in
the perpendicular direction. The opposite situation is found in
other bonding states localized in the system forbidden gap and in
particular for the LU state (see Fig.~\ref{figLUMO}) whose charge
density pertains to the dislocation core laying along the [110]
direction. Clearly, these states are true interface states,
resulting from the reconstructed dislocation cores. It is worth
noting that they would not be obtained from a coherent interface
calculation, and that the determination of the electronic
structure requires the atomic characterization of the misfit
dislocation network.

It has been recently pointed out that a peak is observed in
EELS and XPS around 0.8 eV.\cite{private}
Our results on the presence of interface states in the
heterostructure forbidden gap may give an explanation for this
experimental evidence.

\section{Conclusion}

In conclusion, we have characterized the energetics, the atomic
and electronic structure
of the SiC/Si(001) interface. We performed first principles
DFT-LDA, classical potential and elasticity theory calculations.
The most stable atomic configuration is in agreement with
experiments \cite{Wah94APL,Lon99JAP,Kit00TSF}, where an array of
misfit dislocations pinned at the interface is observed. Actually,
from our results, we also predict that the dislocation core
is characterized by a substoichiometry in carbon atoms. Additional
experiments are needed to confirm our proposed configuration: our
relaxed structures may be used as input for simulating HREM
experimental images and complement the results.

We furthermore estimated the interface energy from ab initio
calculations, for a non-coherent interface, in the case of a
multicomponent system, like SiC: this quantity, which is hardly
accessible from experiment, has been here evaluated for the
SiC/Si(001) interface, although with a non-negligible uncertainty.
This is to our knowledge the first ab initio determination of
interface energies at a mismatched semiconductor heterostructure.

Several electronic interface states, calculated at $\Gamma$, have
been identified. These states, located on the core of the misfit
dislocations, may influence the electronic and optical properties
of the interface.

We point out that the approach we used in this study, {\sl i.e.}
the combination of elasticity theory, classical potential, and ab
initio methods, may be easily adapted for other systems of
interest, in particular for systems with large mismatch, where a
coherent interface model is not suited, or for semiconducting
systems, where core reconstructions are expected.

\begin{acknowledgments}
The authors want to thank Alexis~Baratoff, Pierre~Beauchamp and
Matthieu~George for fruitful discussions. One of us (LP)
acknowledges INFM support for his stay in Italy as visiting
professor. GC acknowledges Demichelis Foundation for financial
support.
Calculations have been performed in CINECA (Bologna, I) through the
INFM Parallel Computing Initiative, at CSCS (Manno, CH), and
at IDRIS (Orsay, F). This work was partially supported by
INFM-PRA:1MESS.
\end{acknowledgments}

\newpage


\begin{thebibliography}{29}
\bibitem[{coh()}]{cohelast}
\bibinfo{note}{The elastic energy per unit area stored in the epilayers due to
  homogeneous strain is $E =
  2\mu\left(\frac{1+\nu}{1-\nu}\right)hf_m^{\,2}$.\cite{Jai97PMA} Taking
  average values for the shear modulus of elasticity $\mu$ and the Poisson's
  ratio $\nu$, and given that the misfit $f_m$ is 0.2 and a single SiC layer
  height $h$ is 1.0875~\AA, we obtained $E\approx1.6$~J/m$^{2}$/layer}.
\bibitem[{not()}]{notePRL}
\bibinfo{note}{Note that these values are very slightly different from those
  given in the reference~\cite{Cic02PRL}, due to a small energy gain obtained
  from a recent improvement in the relaxation of the S1a configuration.}
\end{thebibliography}

\clearpage

\section*{Table captions}

\begin{table}[h]
\caption{Optimized chemical potentials and lattice constants for
Si and 3C-SiC bulks, computed with the Tersoff potential and the
ab initio method. For ab initio, the bulk calculations (column
{\sl ''bulk''}) and the value obtained from linear
extrapolation\protect\cite{Fio96JPCM} (column {\sl ''lin''}) are
indicated for comparison.}\label{chempot}
\end{table}

\begin{table}[h]
\caption{Configuration energy differences, from classical
dynamics, and variation of C and Si atoms, for the first set of
configurations and a 12/8 system (see Fig.~\ref{config1}), with
the geometry S1a taken as the reference.}\label{Energy1}
\end{table}

\begin{table}[h]
\caption{Configuration energy differences for CSS
compared to S1a, within different conditions, and
slab size. As indicated in the text, numbers for the classical
simulations are slightly different from those reported in
Ref.\protect\cite{Cic02PRL}, because of improved convergence. The
ab initio values are obtained with the linear extrapolation
method\protect\cite{Fio96JPCM}, and the consistently derived
chemical potentials were used.}\label{Energy2}
\end{table}

\begin{table}[h]
\caption{Surface energies for the systems used in the slab
calculations, computed with the ab
initio method. The pseudo-hydrogen
energy is included, as detailed in the text.}\label{surfnrj}
\end{table}

\begin{table}[h]
\caption{Valence band Width (VBW) and energy gap ($\Delta_{gap}$)
of the relaxed structures in eV.}\label{VBW}
\end{table}

\clearpage

\vspace{1cm}

\begin{table}[h]
\begin{tabular}{c|cc|ccc}
\hline \hline
 & \multicolumn{2}{c}{Tersoff} \vline & \multicolumn{3}{c}{Ab initio} \\
  & $\mu_0$ (eV) & $a_0$ (\AA)  & $\mu_0$(bulk) (Ha) & $\mu_0$(lin) (Ha) & $a_0$ (\AA) \\
\hline
Si  &    -4.630 &   5.432 & -3.96611 & -3.96877 & 5.401 \\
SiC &   -12.374 &   4.318 & -9.64469  & -9.65374 & 4.334  \\
\hline \hline
\end{tabular}
\end{table}

\vfill
\centerline{\bf Table~\ref{chempot}}

\vspace{1cm}

\begin{table}[h]
\begin{tabular}{c|c|c|cc}
\hline
\hline
Config. & $\Delta n^\mathrm{Si}$ & $\Delta n^\mathrm{C}$ & \multicolumn{2}{c}{$\Delta E_\alpha$~(eV/cell)} \\
  &  &  & C-rich & C-poor \\
\hline
S1a & 0 & 0 & 0.0 & 0.0 \\
   S1b &   0 &   0 & 6.55 & 6.55 \\
 P1 &   0 &  -9 & 16.40 & 9.92 \\
 P2 &  -9 &  -9 & 40.09 & 40.09 \\
  S2 &   0 &   0 & 37.91 & 37.91 \\
C2 & -25 &  16 & 63.74 & 93.26 \\
C1 & -25 &  25 & 63.88 & 99.88 \\
\hline
\hline
\end{tabular}
\end{table}

\vfill
\centerline{\bf Table~\ref{Energy1}}

\clearpage

\vspace{1cm}

\begin{table}[h]
\begin{tabular}{c|cc|cc}
\hline
\hline
$\Delta E_\alpha$ & \multicolumn{2}{c}{Tersoff} \vline & \multicolumn{2}{c}{Ab initio} \\
 (eV/cell) & 12/8 & 36/36 & 5/5 & 7/7 \\
\hline
C-rich & -6.04 & -6.03 & -5.90 & -5.69\\
C-poor & -15.40 & -15.39 &  -15.26  & -15.05 \\
\hline
\hline
\end{tabular}
\end{table}

\vfill
\centerline{\bf Table~\ref{Energy2}}

\clearpage

\vspace{1cm}

\begin{table}[h]
\begin{tabular}{c|cc}
\hline \hline
$E_S$ (Ha/at-H$_2$) &  C-rich & C-poor  \\
\hline
Si(001)-(1x1)  &      \multicolumn{2}{c}{-1.12208}  \\
C:SiC(001)-(1x1)  &      -1.14212 & -1.12890  \\
\hline \hline
\end{tabular}
\end{table}

\vfill
\centerline{\bf Table~\ref{surfnrj}}

\clearpage

\vspace{1cm}

\begin{table}[h]
\begin{tabular}{l |c c c| c}
\hline
\hline
                  &  \multicolumn{3}{c}{VBW (eV)} \vline & $\Delta_{gap}(eV)$ \\
System            &  SiC  &  Si & Total  &   \\
\hline
Si bulk           &    -   & 12.08 &  -   &    -   \\
SiC bulk          &  15.68 &   -   &   -  &    -   \\
5/5 frozen surf.  &  14.67 & 10.91 & 15.31&  0.50  \\
5/5 free surf.    &  14.65 & 9.95  & 14.68&  0.96  \\
7/7 frozen surf.  &  15.06 & 11.25 & 15.57 &  0.40 \\ \hline
\hline
\end{tabular}
\end{table}

\vfill
\centerline{\bf Table~\ref{VBW}}

\clearpage

\section*{Figure captions}

\begin{figure}[h]
\caption{First set of initial configurations. White (black)
circles represent the Si (C) atoms. The dashed line shows the
location of the interface.}\label{config1}
\end{figure}

\begin{figure}[h]
\caption{$[110]$ (left) and $[1\bar{1}0]$ (right) side views of
three relaxed SiC/Si(001) interface configurations. Light grey
(black) spheres show silicon (carbon) atoms. The dashed thick
lines mark the location of the extra atomic planes introduced by
the misfit edge dislocations. Note that the represented bonds are
drawn solely on the basis of a distance criterion and are not
necessarily indicative of a true chemical bond. Removal of the C
atoms inside the ellipses leads to the CSS
configuration.}\label{configrelax1}
\end{figure}

\begin{figure}[h]
\caption{$[110]$ (top) and $[1\bar{1}0]$ (bottom) perspective side
views of the most stable CSS SiC/Si(001) interface configuration.
Light grey (black) spheres show silicon (carbon)
atoms.}\label{configrelax2}
\end{figure}

\begin{figure}[h]
\caption{Layer warping for a 7/7 SiC/Si(001) interface. In
abscissa, the mean value of the coordinate normal to the
interface plane for a given atomic layer is
indicated.}\label{warping}
\end{figure}

\begin{figure}[h]
\caption{Contour plot of the strain field at the 7/7
SiC/Si(001) interface projected along the $[1\bar{1}0]$ direction
for both the CSS (top) and S1a (bottom) geometries, evaluated in
terms of atomic displacements from ideal bulk-like positions. To
enhance comparison, the displacement integrated over a supercell
plane of fixed height from the interface is indicated. Two
lateral replicas are indicated for both configurations, with the
SiC part higher and Si lower; brighter regions depict larger
distortions: they occur slightly below the interface, at the Si
first layer.}\label{stress}
\end{figure}

\begin{figure}[h]
\caption{Interaction energy between surface and interface as a
function of the slab size, for the S1a (left) and CSS (right)
configurations.}\label{surfinternrj}
\end{figure}

\begin{figure}[h]
\caption{Calculated DOS for the CSS configuration and a 7/7 slab,
projected on the Si and SiC layers at the interface (top panel),
compared with the DOS for bulk SiC and Si (middle and bottom
panels). The width of Si and SiC bands, as well as the energies of
HO and LU states, are shown in the figure.}\label{CSSdos}
\end{figure}

\begin{figure}[h]
\caption{Isosurface (medium grey) of the highest
occupied state projected along $[110]$ (left) and $[1\bar{1}0]$
(right). Black (light grey) spheres indicate C (Si) species.
Surface atoms are saturated with hydrogens (small white
spheres).}\label{figHOMO}
\end{figure}

\begin{figure}[h]
\caption{Isosurface (medium grey) of the lowest unoccupied
state projected along $[110]$ (left) and $[1\bar{1}0]$ (right).
Black (light grey) spheres indicate C (Si) species. Surface atoms
are saturated with hydrogens (small white
spheres).}\label{figLUMO}
\end{figure}

\clearpage

\ \vspace{2cm}

\begin{center}
\includegraphics[width=16cm]{fig1.eps}
\end{center}

\vfill\centerline{\bf Figure~\ref{config1}}

\clearpage

\ \vspace{2cm}

\begin{center}
\includegraphics[width=14cm]{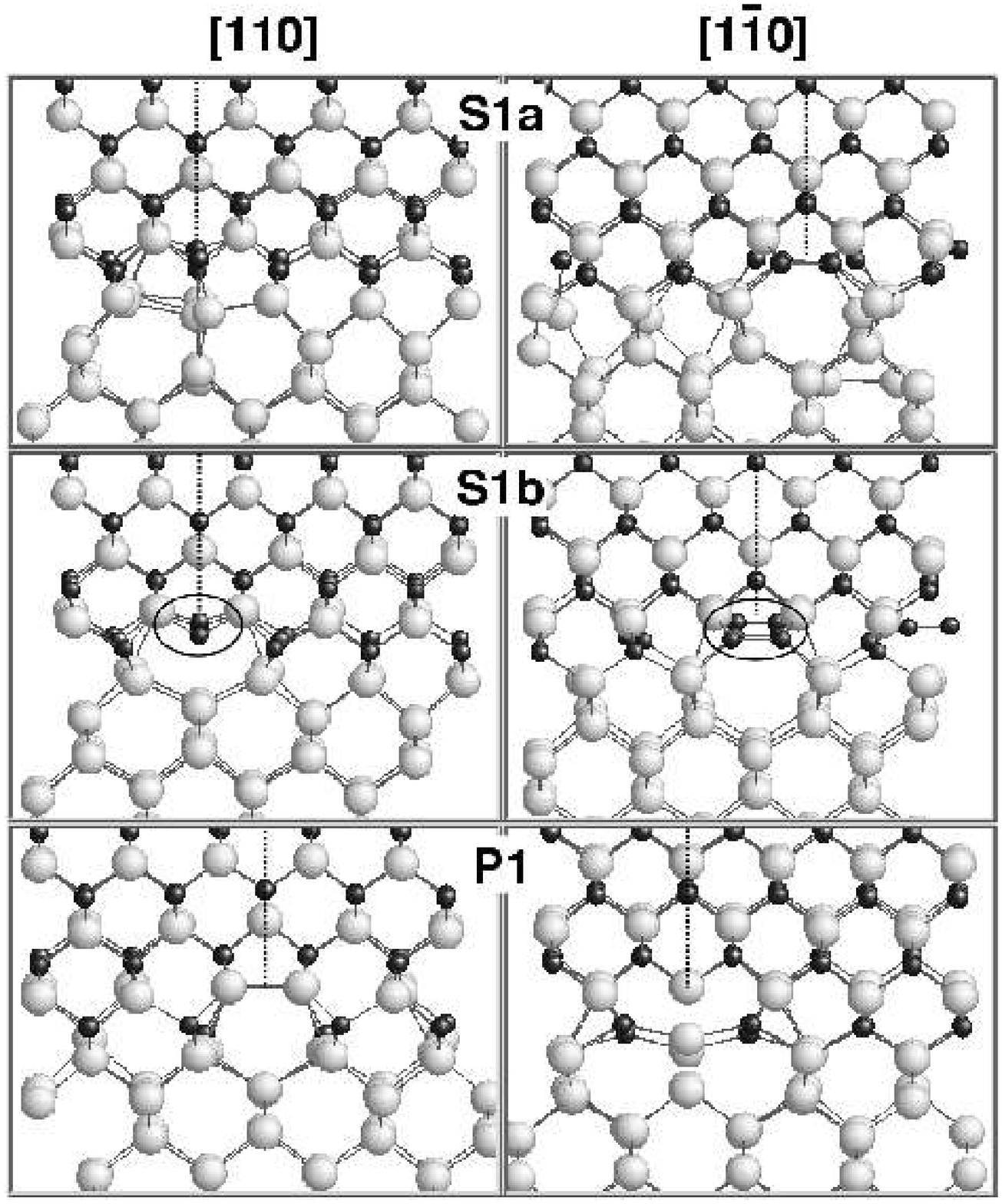}
\end{center}

\vfill\centerline{\bf Figure~\ref{configrelax1}}

\clearpage

\ \vspace{2cm}

\begin{center}
\includegraphics[width=14cm]{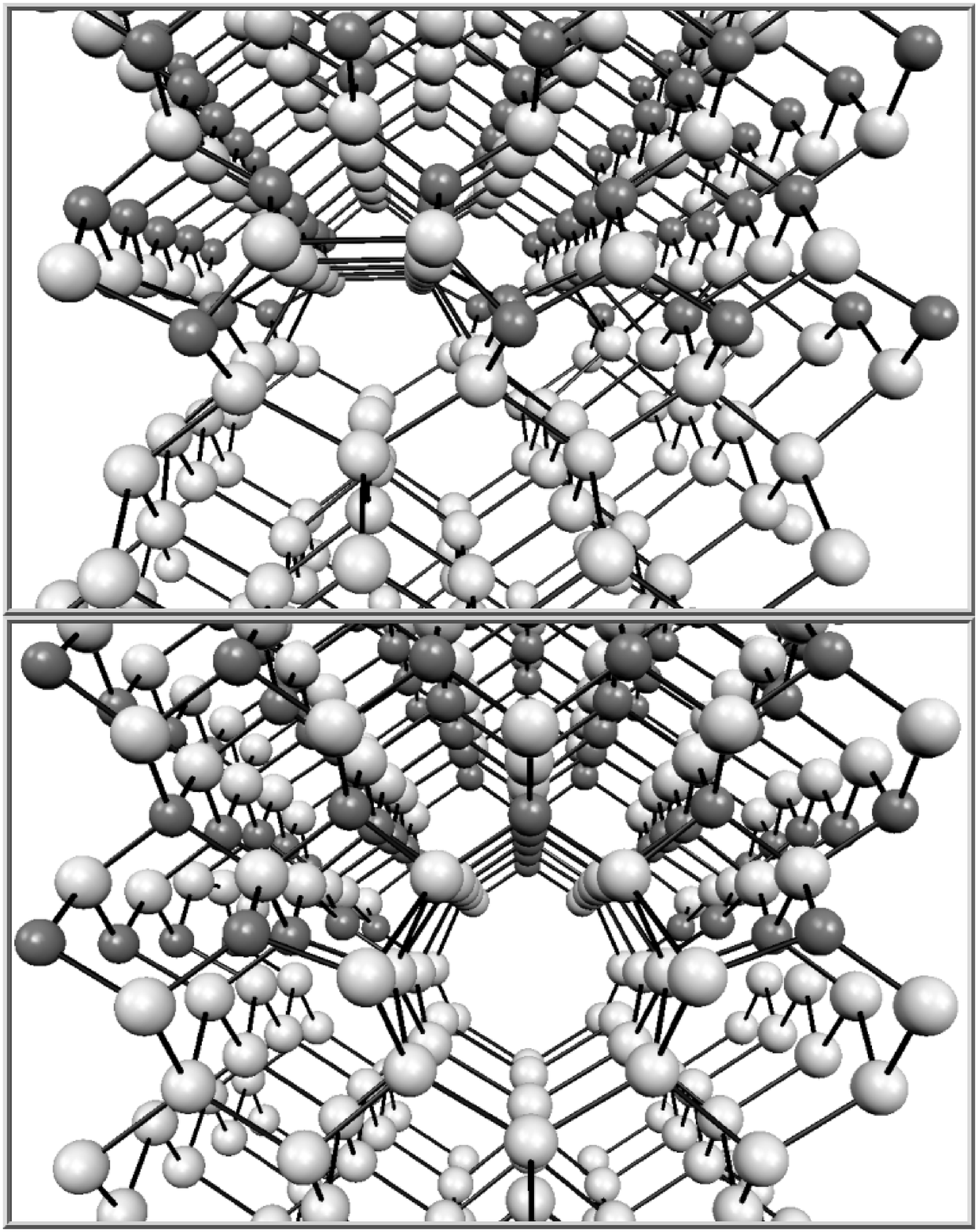}
\end{center}

\vfill\centerline{\bf Figure~\ref{configrelax2}}

\clearpage

\ \vspace{2cm}

\begin{center}
\includegraphics[width=16cm]{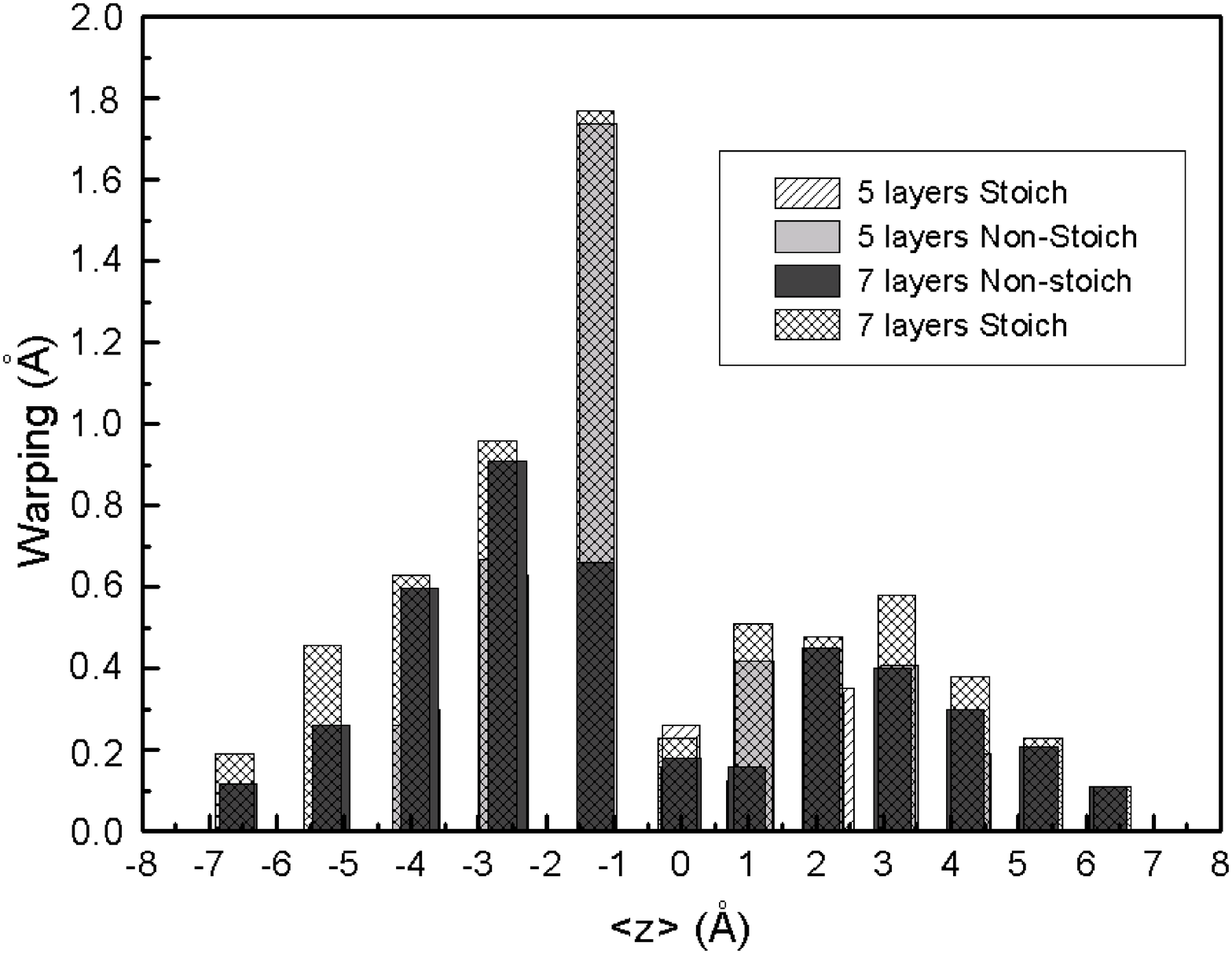}
\end{center}

\vfill\centerline{\bf Figure~\ref{warping}}

\clearpage

\ \vspace{2cm}

\begin{center}
\includegraphics[width=16cm]{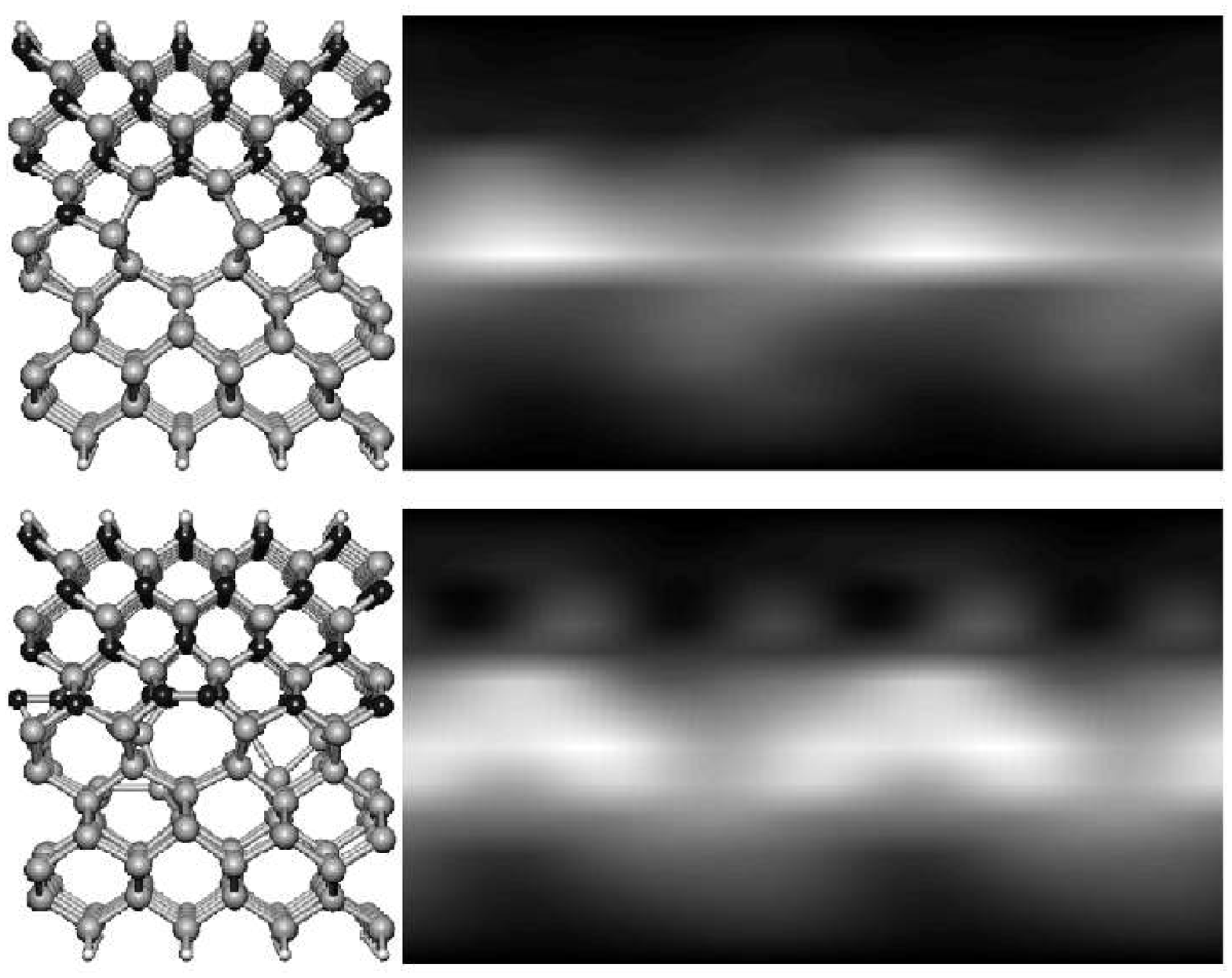}
\end{center}

\vfill\centerline{\bf Figure~\ref{stress}}

\clearpage

\ \vspace{2cm}

\begin{center}
\includegraphics[width=16cm]{fig6.eps}
\end{center}

\vfill\centerline{\bf Figure~\ref{surfinternrj}}

\clearpage

\vspace{2cm}

\begin{center}
\includegraphics[width=16cm]{fig7.eps}
\end{center}

\vfill\centerline{\bf Figure~\ref{CSSdos}}

\clearpage

\ \vspace{2cm}

\begin{center}
\includegraphics[width=18cm]{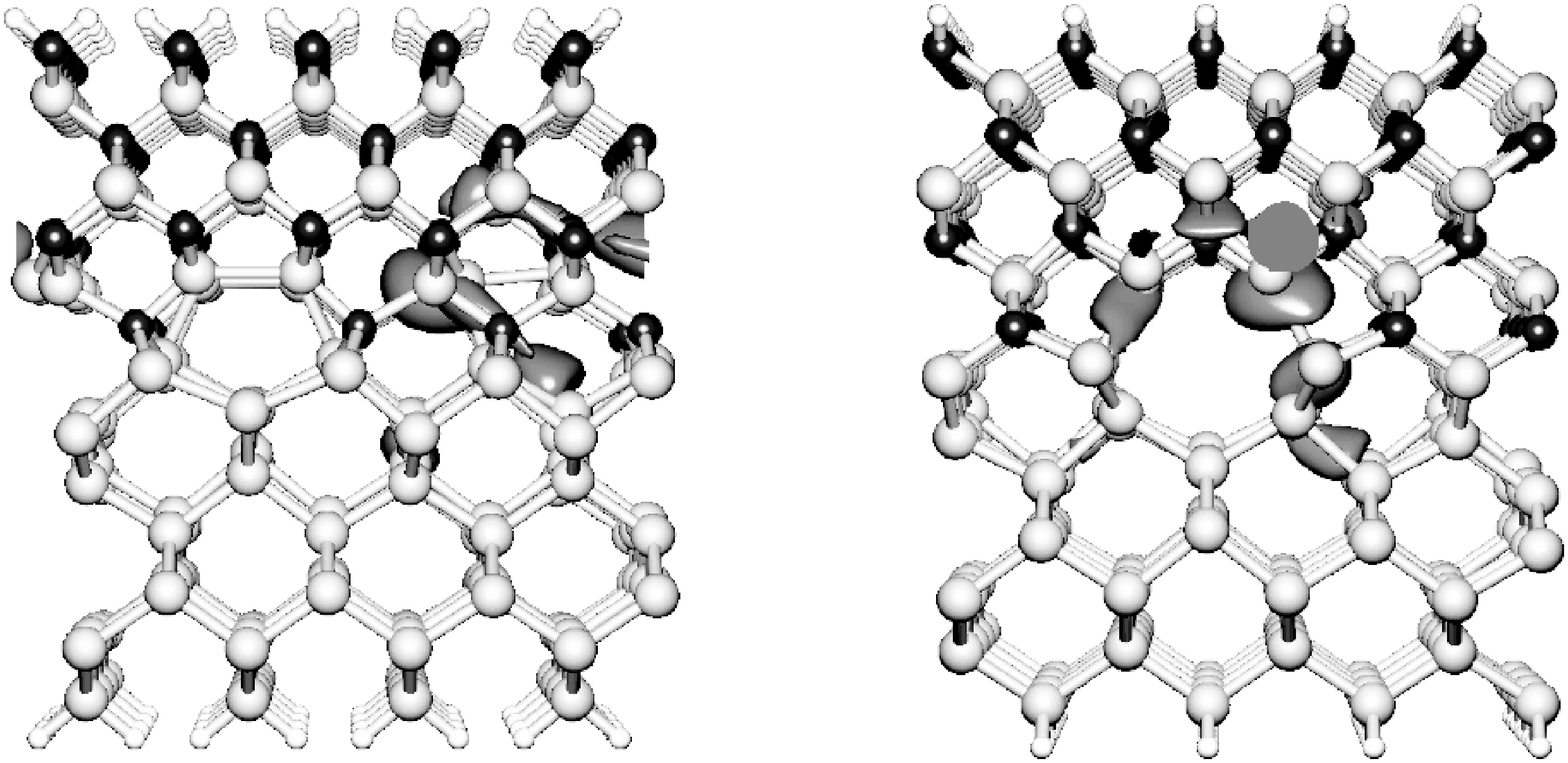}
\end{center}

\vfill\centerline{\bf Figure~\ref{figHOMO}}

\clearpage

\ \vspace{2cm}

\begin{center}
\includegraphics[width=18cm]{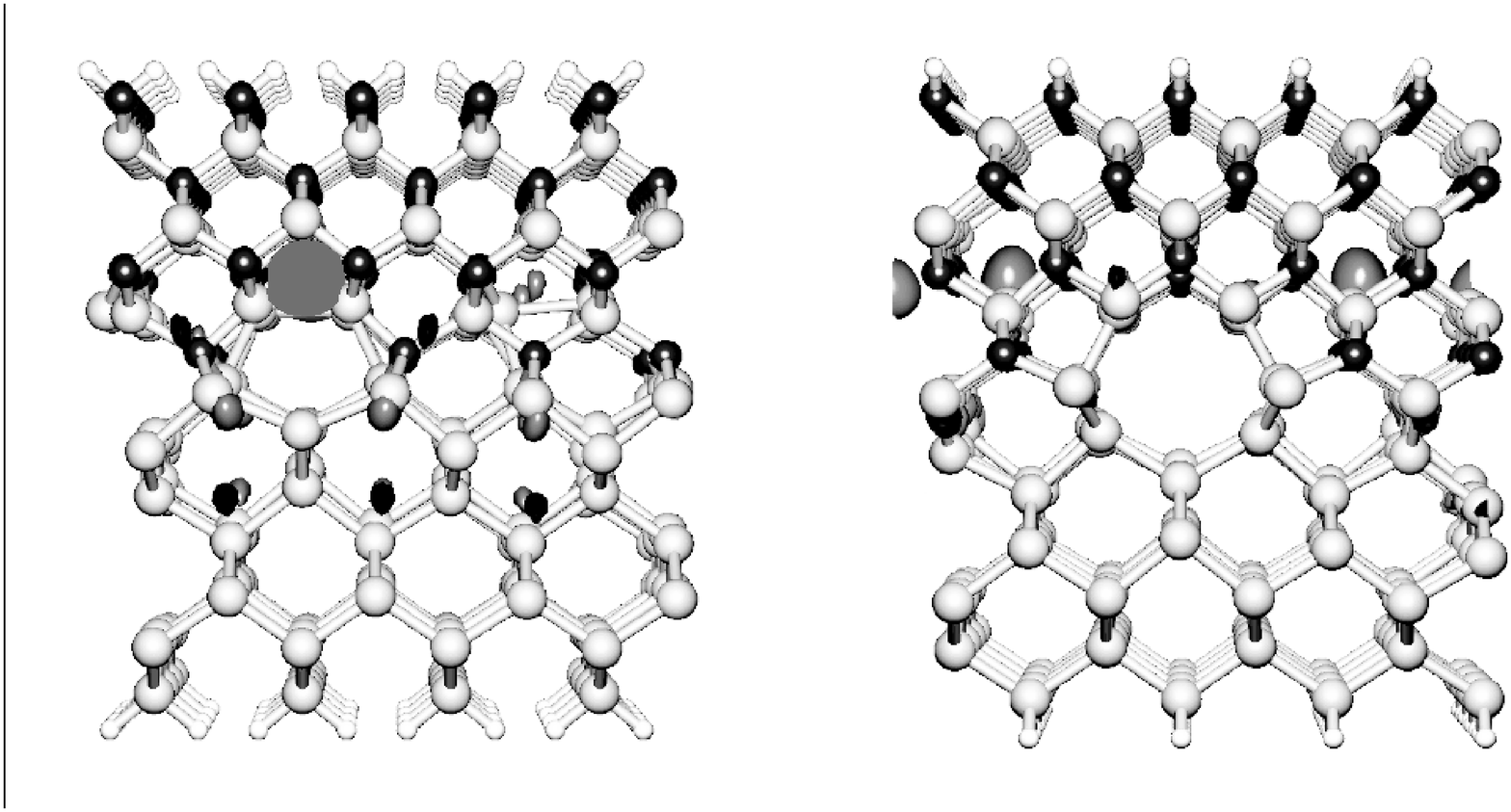}
\end{center}

\vfill\centerline{\bf Figure~\ref{figLUMO}}

\end{document}